\pdfoutput=1

\documentclass[sigconf,techreport]{acmart}

\makeatletter
\def\mdseries@tt{m}
\makeatother

\usepackage{fvextra}
\fvset{frame=single,
       samepage=true,
       fontfamily=courier,
       numbers=left,
       numbersep=2pt,
       baselinestretch=0.90,
       fontsize=\scriptsize}

\everypar{\looseness=-1}

\makeatletter
\@ifclasswith{acmart}{techreport}{
  \settopmatter{printacmref=false}
  \renewcommand\footnotetextcopyrightpermission[1]{}
  \fancyhead{}
  \fancyfoot{}
  \fancyhead[L]{\shorttitle}
  \fancyhead[R]{\shortauthors}
  \fancyfoot[C]{\thepage}
}{}
\makeatother

\begin{document}

\title{Impact of HTTP Cookie Violations in Web Archives}

\author{Sawood Alam, Plinio Vargas, Michele C. Weigle, and Michael L. Nelson}
\affiliation{%
  \institution{Department of Computer Science, Old Dominion University, Norfolk, Virginia -- 23529 (USA)}
}
\email{{salam,pvargas,mweigle,mln}@cs.odu.edu}

\renewcommand{\shortauthors}{S. Alam et al.}

\begin{abstract}
Certain \emph{HTTP Cookies} on certain sites can be a source of content bias in archival crawls.
Accommodating \emph{Cookies} at crawl time, but not utilizing them at replay time may cause cookie violations, resulting in defaced composite mementos that never existed on the live web.
To address these issues, we propose that crawlers store \emph{Cookies} with short expiration time and archival replay systems account for values in the \texttt{Vary} header along with URIs.
\end{abstract}

\maketitle

\section{Introduction and Background}

\begin{figure}
  \includegraphics[width=0.82\linewidth]{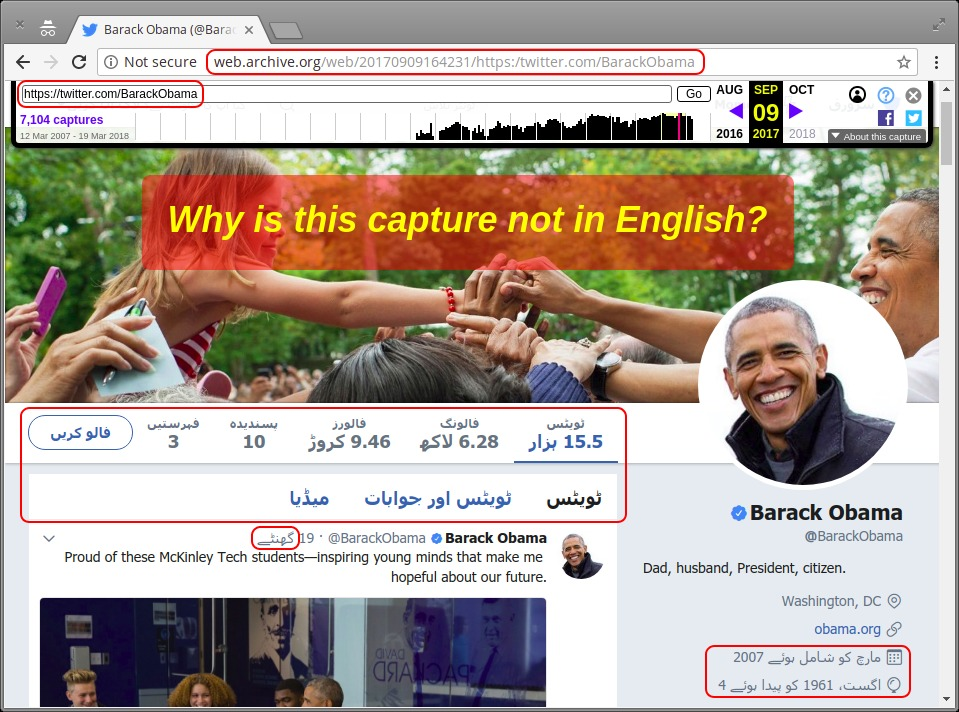}
  \vspace{-0.4cm}
  \caption{Barack Obama's Twitter Timeline is Archived in Urdu, Which Should be in English}
  \label{img:barackobama-twitter-urdu-annotated}
\end{figure}

\begin{figure}
  \includegraphics[width=0.82\linewidth]{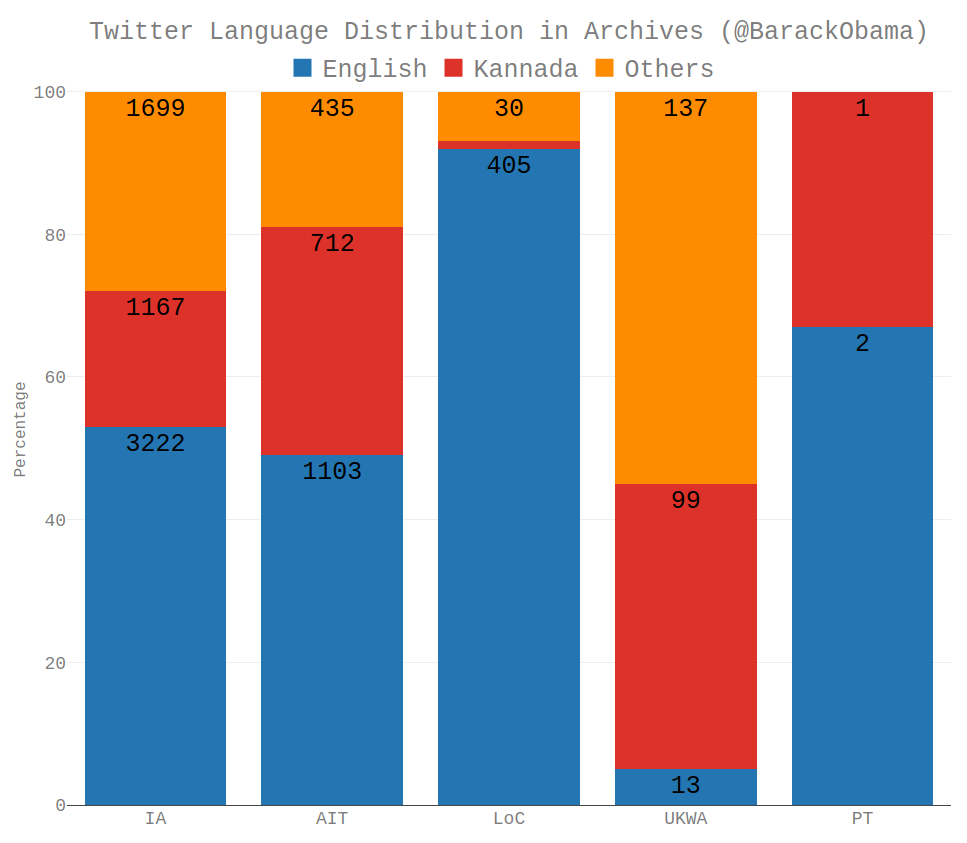}
  \vspace{-0.4cm}
  \caption{Language Distribution of Mementos of Barack Obama's Twitter Timeline in Different Web Archives}
  \label{img:barackobama-lang-dist}
\end{figure}

For a long time we have been observing a strange behavior of various web archives when accessing mementos~\cite{memento:rfc} of Twitter pages, some of the mementos would be replayed in non-English languages.
This happens even if those Twitter timelines belong to English-speaking personalities, archived using crawlers in North America, and were not requested in any specific language explicitly as shown in Figure~\ref{img:barackobama-twitter-urdu-annotated}.
After a thorough investigation we figured it out that it is happening due to the use of \emph{HTTP Cookies} for content negotiation by Twitter~\cite{cookie-lang:blog,dshr-kannada:blog}.
We found that almost half of the mementos of Barack Obama's Twitter timeline out of over 9,000 properly archived mementos in five different web archives were in non-English languages, of which, almost half were in Kannada (a regional Indian language) alone, and remaining in 45 other languages (as shown in Figure~\ref{img:barackobama-lang-dist}).
While language diversity in web archives is generally a good thing, this non-uniform bias is disconcerting when a page is archived in a language not anticipated.

\begin{figure}
  \includegraphics[width=0.82\linewidth]{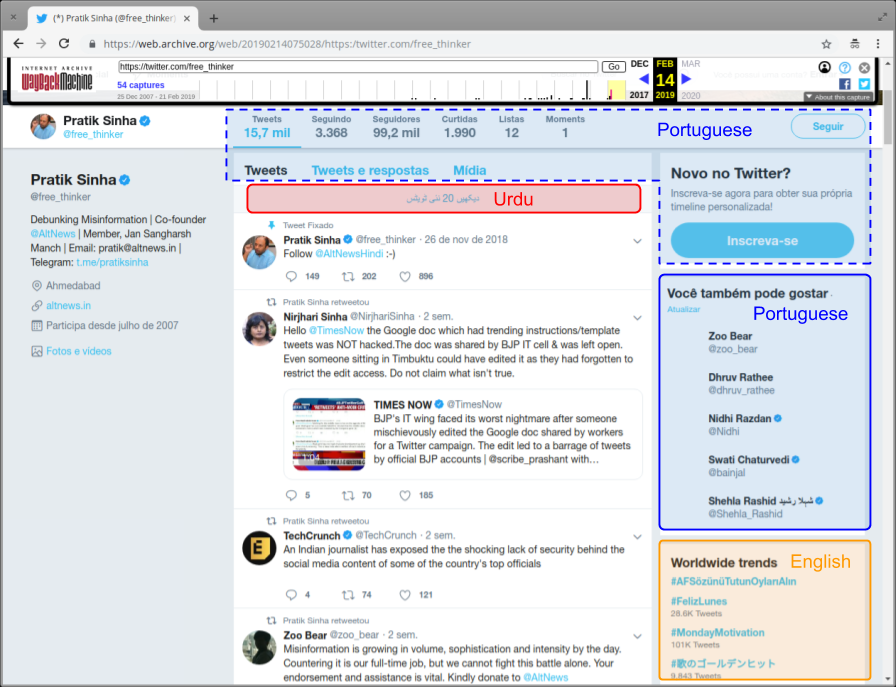}
  \vspace{-0.4cm}
  \caption{A Memento of a Twitter Timeline Simultaneously in Multiple Languages (Portuguese, English, and Urdu)}
  \label{img:twitter-language-mixed-annotated-highlighted}
  \vspace{-0.5cm}
\end{figure}

One day we were looking at a Twitter timeline's memento which should have been in English, but was primarily in Portuguese (for the reason described above), after a while we noticed that a notification appeared in Urdu, suggesting that there were 20 new tweets (as shown in Figure~\ref{img:twitter-language-mixed-annotated-highlighted}).
On further inspection found that the page contained a sidebar block in English too.
Apparently, we were seeing a defaced composite memento of a page that perhaps never existed on the live web.
We knew that live-leakage (also known as Zombies)~\cite{zombies:blog} and temporal violations~\cite{hypertext-2015-ainsworth-emporal-violations} can cause such malformed memento reconstruction and we also knew their potential prevention techniques~\cite{sigsac-2017-lerner-rewriting,jcdl-2017-alam-reconstructive}.
However, this mixed-language Twitter timeline issue cannot be explained by zombies nor temporal violations.
After a thorough investigation we found that \emph{Cookies} were again the reason behind this replay issue~\cite{cookie-violations:blog,dshr-47links:blog}.

\emph{HTTP} is stateless, but often applications need to maintain state information between a client and a server.
This is often done with the help of \emph{Cookies}~\cite{cookie:rfc}.
Servers can send one or more \texttt{Set-Cookie} headers containing strings of name-value pairs along with scope (domain and path) and expiration information.
Clients store them and send them back with each request in the scope using \texttt{Cookie} header until expired or removed.
\emph{Cookies} are used for session management, personalization, content-negotiation, tracking, and client-side key-value store.
The latter is less common now after wide adoption of \emph{LocalStorage} and other similar techniques in web browsers.

\section{Internationalization in Twitter}

Twitter uses standard method of internationalization in its publicly accessible pages by including \texttt{alternate} links in 47 supported languages and the \texttt{x-default} landing language (as illustrated in Figure~\ref{code:twitter-alt-lang}) to help search engines point users with different locales to the correct language.
This technique is utilized by many other multi-lingual sites such as Facebook and Instagram.
However, unlike other popular multi-lingual sites, when accessing a language-specific URI (that contains a \texttt{lang} query parameter), Twitter sets a \texttt{lang} \emph{Cookie} with the corresponding language (as illustrated in Figure~\ref{code:twitter-lang-cookie}).
This \emph{Cookie} sticks throughout the session and forces all subsequent pages to be served in that language until another language-specific URI overwrites the \emph{Cookie}.
This essentially means Twitter performs language content negotiation using \emph{Cookie} header, though it does not acknowledge it in a \texttt{Vary} header.

\begin{figure}
\begin{Verbatim}[commandchars=^\{\}]
<link rel="alternate" hreflang="x-default"
                      href="https://twitter.com/">
<link rel="alternate" ^textcolor{red}{hreflang="fr"}
                      href="https://twitter.com/?^textcolor{red}{lang=fr}">
... ^textit{[45 LINKS TRUNCATED]} ...
<link rel="alternate" ^textcolor{red}{hreflang="kn"}
                      href="https://twitter.com/?^textcolor{red}{lang=kn}">
\end{Verbatim}
\vspace{-0.4cm}
\caption{47 Alternate Language Links in Twitter}
\label{code:twitter-alt-lang}
\vspace{-0.5cm}
\end{figure}

\begin{figure}
\begin{Verbatim}[commandchars=^\{\}]
^textbf{$ curl -s ^textcolor{red}{-c /tmp/tt.cook} https://twitter.com/?^textcolor{red}{lang=ar} \}
^textbf{>      | grep "<html"}
<html ^textcolor{red}{lang="ar"} data-scribe-reduced-action-queue="true">
^textbf{$ grep ^textcolor{red}{lang} /tmp/tt.cook}
twitter.com FALSE / FALSE 0 ^textcolor{red}{lang ar}
^textbf{$ curl -s https://twitter.com/ | grep "<html"}
<html ^textcolor{red}{lang="en"} data-scribe-reduced-action-queue="true">
^textbf{$ curl -s ^textcolor{red}{-H "Accept-Language: ur"} https://twitter.com/ \}
^textbf{>      | grep "<html"}
<html ^textcolor{red}{lang="ur"} data-scribe-reduced-action-queue="true">
^textbf{$ curl -s ^textcolor{red}{-b /tmp/tt.cook} https://twitter.com/ | grep "<html"}
<html ^textcolor{red}{lang="ar"} data-scribe-reduced-action-queue="true">
\end{Verbatim}
\vspace{-0.4cm}
\caption{Language Content Negotiation in Twitter Using Query Parameters, Accept-Language, and Cookies}
\label{code:twitter-lang-cookie}
\vspace{-0.5cm}
\end{figure}

\section{Cookie Violations}

Some websites insist that certain \emph{Cookies} are present in a request before they return desired content otherwise they issue redirects and attempt to set those \emph{Cookies}.
Failure to send their desired \emph{Cookies} in subsequent requests may turn such sites into crawler traps without any useful content.
Web archiving crawlers such as Heritrix\footnote{\url{https://github.com/internetarchive/heritrix3}} have built-in support for cookies.
However, the web surfing pattern of crawlers is generally breadth-first-style and comprehensive (not necessarily how human surf the web) for which they use frontier queue of URIs to be crawled.
In case of the Twitter's example above, when one of the language-specific \texttt{alternate} link is crawled, it impacts all the subsequent non-language-specific URIs due to the \texttt{lang} sticky \emph{Cookie}.
Kannada being the last language in the list (in Figure~\ref{code:twitter-alt-lang}) gets more exposure before it gets overwritten by another language, resulting in the disproportionate language bias.

Popular archival replay systems (such as OpenWayback\footnote{\url{https://github.com/iipc/openwayback}} and PyWB\footnote{\url{https://github.com/webrecorder/pywb}}) utilize only the canonicalized \emph{URI-R} and the datetime of the capture to select a memento to replay.
Other request headers that might have been used for content negotiation (such as \texttt{Accept-Language} or \texttt{Geolocation} etc.) are ignored at replay.
Traditional crawlers did not execute JavaScript, so the likelihood of a custom request header being utilized during crawling was minimal, but it is changing with headless browser-based crawlers.
\emph{Cookies}, however, have been supported even in traditional crawlers that are used by some sites for content negotiation (as is the case with Twitter).
Moreover, aggregating private archives~\cite{jcdl-2018-mkelly-private-archive} containing authenticated resources without isolating them based on session \emph{Cookies} has some privacy implications.

Based on our assessment we propose that \emph{Cookies} in crawlers are kept short-lived and pruned frequently to minimize the impact of sticky \emph{Cookies}.
Accommodating \emph{Cookies} (or other headers that affect the response) at capture/crawl time, but not utilizing them at replay time has this consequence of cookie violations, resulting in defaced composite mementos.
On the contrary, blindly utilizing every \emph{Cookie} as a filter at replay would result in many false negatives.
Unfortunately, \emph{Cookie} names are opaque strings and carry no agreed upon semantics to identify ones that affect the payload.

\vspace{-0.15cm}

\section{Conclusions and Future Work}

We identified that certain \emph{Cookies} on certain sites can be a source of content bias in archival crawls.
To address this issue we propose that crawlers store \emph{Cookies} with short expiration time explicitly, irrespective of the original value.
We also identified that \emph{Cookie Violations} at replay time have the potential to deface composite mementos and reconstruct pages from web archives that never existed on the live web.
Archival replay systems need to behave like HTTP proxies or cache servers that accommodate values in the \texttt{Vary} header along with URIs.
Not every \emph{Cookie} is created equal, those impacting the content need to be identified and accounted for at replay.
This is a difficult problem which opens up the possibility for a more extensive research to fully address the issue.

\vspace{-0.15cm}

\section{Acknowledgements}

This work is supported in part by NSF grant IIS-1526700.

\vspace{-0.15cm}

\bibliographystyle{ACM-Reference-Format}
\bibliography{ref} 

\end{document}